\documentclass{article}
\input amssym.def
\input amssym.tex
\usepackage[a4paper, left=2cm,top=2.5cm,right=2cm,bottom=3cm,nohead]{geometry}
\def\ben{\begin{equation}}
\def\een{\end{equation}}
\def\half{\frac{1}{2}}
\def\bea{\begin{eqnarray}}
\def\eea{\end{eqnarray}}

\def\bx{{\bf x}}

\def\p{\partial}
\def\mathbb{\Bbb}

\def\ben{\begin{equation}}
\def\een{\end{equation}}
\def\half{{1 \over 2}}
\def\bea{\begin{eqnarray}}
\def\eea{\end{eqnarray}}

\def \p{\partial}
\def\p{\partial}
 
\def\bom{{\mbox{\boldmath $ \omega $ }}}

\def\bx{{\bf x}}
\def\bM{{\bf M}}
\def\bom{{\mbox{\boldmath $ \omega $ }}} 
\def\bk{{\bf k}}
  \def\p{{\partial}} 
\newcount\hour \newcount\minute
\hour=\time  \divide \hour by 60
\minute=\time
\loop \ifnum \minute > 59 \advance \minute by -60 \repeat
\def\nowtwelve{\ifnum \hour<13 \number\hour:
                      \ifnum \minute<10 0\fi
                      \number\minute
                      \ifnum \hour<12 \ A.M.\else \ P.M.\fi
         \else \advance \hour by -12 \number\hour:
                      \ifnum \minute<10 0\fi
                      \number\minute \ P.M.\fi}
\def\nowtwentyfour{\ifnum \hour<10 0\fi
                \number\hour:
                \ifnum \minute<10 0\fi
                \number\minute}

\title{Some Spacetimes with Higher Rank Killing--St\"ackel Tensors}
\author{G.~W.~Gibbons$^1$, T.~Houri$^2$, D.~Kubiz{\v n}\'ak$^1$ and 
  C.~M.~Warnick$^{1, 3}$ 
\\
\\ \small{1. D.A.M.T.P., Cambridge, Wilberforce Road, Cambridge CB3 0WA,
  U.K.}
\\ \small{2.  Osaka City University Advanced Mathematical Institute (OCAMI)}
\\ \small{ 3-3-138 Sugimoto, Sumiyoshi, Osaka 558-8585, Japan }   
\\ \small{3. Queens' College, Cambridge, CB3 9ET, U.K.} 
\\}

\begin{document}

\maketitle {\let\thefootnote\relax\footnotetext{{\em Emails}:
  gwg1@damtp.cam.ac.uk, cmw50@damtp.cam.ac.uk, dk317@damtp.cam.ac.uk
, houri@sci.osaka-cu.ac.jp \\ \mbox{} \hspace{.45cm}\emph{Pre-print no.}
 DAMTP-2011-20, OCU-PHYS 349}}

\begin{abstract}
By applying the lightlike Eisenhart lift to several known examples of
low-dimensional integrable systems admitting 
integrals of motion of higher-order in momenta, we obtain four- and higher-dimensional Lorentzian spacetimes with irreducible 
higher-rank Killing tensors. Such metrics, we believe, are first examples of spacetimes admitting 
higher-rank Killing tensors. Included in our examples is a
four-dimensional supersymmetric pp-wave spacetime, whose geodesic flow
is superintegrable. The Killing tensors satisfy a non-trivial
Poisson--Schouten--Nijenhuis algebra. We discuss the extension to the
quantum regime.
\end{abstract}



\section{Introduction}

Since Carter's {\it tour de force} in separating variables
for the Hamilton--Jacobi and Klein--Gordon equations in the Kerr metric 
\cite{Carter} there has been a great deal of work on spacetimes
$\{{\cal M},g_{ab}\}$ 
admitting  a second rank Killing--St\"ackel  tensor  $K^{ab}= K^{ba} $ which is 
responsible for the additive separability of the 
 Hamilton--Jacobi equation.  
Almost nothing is known about 
higher rank  totally symmetric tensors 
$K^{a_1a_2\dots a_p}$ satisfying the condition that      
\ben
\nabla ^{(a_1} K^{a_2 a_3\dots a_{p+1} ) } =0 \,. 
\een
 While it is known  that any such tensor gives rise
to a homogeneous function on the cotangent bundle  $T^\star{\cal M}$,  ${\cal K}_p\! =\!K^{a_1\dots a_p}p_{a_1}\dots p_{a_p}$ of degree $p$
  in momenta, which Poisson  commutes
with the Hamiltonian $ {\cal H} =\half g^{ab}p_a p_b$ generating the
geodesic flow,
no non-trivial (i.e. irreducible) examples appear to be known.

Given any two such Killing--St\"ackel  tensors of rank  $p$ and $q$ respectively
their Schouten--Nijenhuis bracket 
$[K_p ,K_q ] ^{a_1a_2 \dots a_{p+q-1} }$ is  defined in terms of the standard Poisson bracket $\{{\cal K}_p, {\cal K} _q \}$  as follows
\ben
\Big \{ {\cal K}_p, {\cal K} _q \Bigr \} = \frac{\partial {\cal K}_p}{\partial q^i} \frac{\partial {\cal K}_q}{\partial {p_i}}-\frac{\partial {\cal K}_q}{\partial q^i} \frac{\partial {\cal K}_p}{\partial {p_i}}\equiv 
[K_p ,K_q ] ^{a_1a_2 \dots a_{p+q-1} } p_{a_1} p_{a_2}
 \dots p_{a_{p+q-1} } \,.
\een

While examples  of  spacetimes
admitting more than one quadratic Killing tensor
satisfying a non-trivial  Poisson or Schouten--Nijenhuis bracket algebra
exist \cite{GDH}, no such higher rank examples appear to be known.
This may well be because the quickest route for 
finding quadratic Killing tensors is to follow Carter's original path \cite{Carter} and seek to separate variables in the 
Hamilton--Jacobi and Klein--Gordon equations. This route is not available
for higher rank Killing--St\"ackel  tensors since there is no obvious
connection between their  existence and separability. 
By theorems in \cite{BenentiFrancaviglia:1979, KalninsMiller:1981}
only rank two Killing tensors apply to separability of the Hamilton--Jacobi equation.
   
In some cases it is possible to  go further and ``quantize''
the system. In the case of quadratic Killing--St\"ackel  tensors
it is known that subject to  certain  conditions  on the $K^{ab} $
and the Ricci tensor
$R_{ab}$  , the second order differential operator
$ -\nabla_a  K^{ab} \nabla_b$
commutes with the wave operator 
$ -\nabla_a  g^{ab} \nabla_b$  and this is related
to the multiplicative separability of the Klein--Gordon equation 
\cite{Carter:1977}. A recent survey of quantum integrability
of quadratic Killing--St\"ackel tensors  may be found in \cite{V}.
To our  knowledge, there are few if any results to date on the higher
rank case. 

The paper is organized as follows. In section 2 we give details of the
lightlike Eisenhart lift and in particular how constants of the motion
are lifted. In section 3 we give examples of spacetimes generated from
the classical examples of Liouville integrable dynamical systems
describing heavy tops. In section 4 we discuss how to obtain a
supersymmetric spacetime by lifting dynamical systems in $E^2$ and
give a superintegrable example. We conclude in section 5 and include a
brief summary of conventions in the appendix.

\section{The Eisenhart Lift}

Our examples are all obtained by taking the {\it Eisenhart lift}
or oxidation
\cite{GDH,Eisenhart,Minguzzi,Gibbons} 
of a dynamical system with an $n$-dimensional configuration
space $\{Q_n, g_{ij}, V, A_i \}  $ with Lagrangian 
\ben
L= \half g_{ij}(q^k,t) \dot q^i \dot q^j  - V(q^k,t) + A_i(q^k,t) \dot q^i \,,  
\een
to  give a system of  geodesics 
in an  $(n+2)$-dimensional {\it Bargmann spacetime}  
$\{{\cal M}, g_{ab}, \partial_s   \} $, which admits  a covariantly constant  null
Killing vector field $\partial_s$. The original dynamical trajectories
are obtained  by a null reduction along the orbits of 
$\partial_s$. 
Since all Bargmann metrics admit a covariantly constant null vector
field, it follows that the holonomy is contained within $E(2) \subset SO(3,1)$, the two-dimensional Euclidean group which stabilizes a null
vector. Thus the null congruence is
geodesic,  expansion, shear and vorticity free.
Thus it is also  contained within the class of Kundt spacetimes.

It is simplest to work with the Hamiltonian formulation in order to see how the lift affects constants of the motion. We consider dynamics on the cotangent bundle, 
$T^*M$, of some manifold
$M$ which is equipped with a natural symplectic form given in local
coordinates by $\omega = dq^i \wedge dp_i$, with associated Poisson
bracket $\{, \}$. We assume that the Hamiltonian is a polynomial of
degree two in momenta:
\begin{equation}
H = H^{(2)} +H^{(1)}+H^{(0)}, \label{Hdef}
\end{equation}
where $H^{(i)}$ has degree $i$ in momenta. We do not need to assume
that $H$ is independent of $t$. We lift $H$ to a Hamiltonian on $T^*(M
\times \mathbb{R}^2)$ by promoting $t$ to a configuration space
coordinate and introducing a new coordinate $s$. The conjugate momenta
are denoted $p_t, p_s$ and the new symplectic form is $\omega' =
\omega + dt \wedge dp_t + ds \wedge dp_s$, with associated Poisson
bracket $\{, \}'$. The Hamiltonian on this
enlarged phase space is
\begin{equation}
\mathcal{H}= H^{(2)} +p_s H^{(1)}+ p_s^2 H^{(0)} + p_s p_t.
\end{equation}
Projecting the integral curves of this system onto the $T^*M \times
\mathbb{R}_t$ factor of the phase space gives integral curves of the
original Hamiltonian. 

Suppose now that the system $(H, T^*M)$ has a constant of the motion
which is a polynomial in momenta:
\begin{equation}
K = \sum_{i=0}^{k} K^{(i)}.
\end{equation}
We calculate the variation of $K$ along an integral curve of $(H,
T^*M)$ and find after collecting terms according to their degree in
momenta that
\begin{equation}
0=\frac{dK}{dt} =\{K, H\}+\frac{\partial K}{\partial t} = \sum_{i=0}^k\left [ \{K^{(i-1)}, H^{(2)}\}+ \{K^{(i)},
  H^{(1)}\}+ \{K^{(i+1)}, H^{(0)}\} +\frac{\partial
  K^{(i)}}{\partial t}\right ],
\end{equation}
Since $K$ should be constant along \emph{any} integral curve, the
terms in the sum should vanish independently for each $i$. We lift $K$
to the extended phase space as
\begin{equation}
\mathcal{K} = \sum_{i=0}^{k} p_s^{k-i} K^{(i)}.
\end{equation}
Now, along an integral curve of $(\mathcal{H},T^*(M\times
\mathbb{R}^2))$ we have
\begin{equation}
 \frac{d\mathcal{K}}{d\lambda} =\{\mathcal{K}, \mathcal{H}\}' =  \sum_{i=0}^k p_s^{k-i+1} \left [ \{K^{(i-1)}, H^{(2)}\}+ \{K^{(i)},
  H^{(1)}\}+ \{K^{(i+1)}, H^{(0)}\} +\frac{\partial
  K^{(i)}}{\partial t}\right ],
\end{equation}
Clearly this vanishes iff $K$ is a constant of the motion for the
original system. Furthermore, since $\mathcal{H}$ is a homogeneous
polynomial of degree two in momenta we may interpret it as generating
the geodesic flow of a (pseudo-)Riemannian metric. $\mathcal{K}$ is a
constant along geodesics which is a homogeneous polynomial in momenta
and so corresponds to a Killing tensor of this metric. A similar
calculation shows that for constants of the motion for the original
system $K_1, K_2, K_3$ which lift to $\mathcal{K}_1, \mathcal{K}_2,
\mathcal{K}_3$ we have
\begin{equation}
\{ K_1, K_2 \} = K_3, \quad \Leftrightarrow \quad \{ \mathcal{K}_1, \mathcal{K}_2 \}' = \mathcal{K}_3.
\end{equation}
As a result, the Shouten--Nijenhuis algebra of the Killing tensors in
the lifted spacetime will be the same as the Poisson algebra of the
constants of the motion for the original dynamical system. We also
note that whilst we have increased the dimension of the configuration
space by two, we have also gained\footnote{%
The equations of motion derived from
$\mathcal{H}$ imply that $p_t = \mbox{const}-E(t)/p_s$, where $E(t)$ is the energy
of the original system, thus when $E$ is constant, we do not lose
this constant of the motion by lifting.
} 
two new constants of the motion: $p_s$ and $p_t$. Thus the \emph{degree of
  integrability} of the system is unchanged by the lift---if the
original system is Liouville integrable (i.e.\ admits $n$ functionally independent constants of the motion
in involution) or super-integrable (admits further constants of
the motion) then so will the lifted system be.

Applying this method to the system $\{Q_n, g_{ij}, V, A_i \}  $ defined above, we find that the lifted system is equivalent to geodesic motion on the spacetime with metric
\ben
ds^2= g_{ij}(q^k,t)  dq^i dq^j  - 2 V(q^k,t)dt^2 + 2 A_i(q^k,t) d q^i dt + 2 dt ds \,.  
\een

\section{Eisenhart lift of Goryachev--Chaplygin and Kovalevskaya's Tops}

\subsection{Eisenhart lift of the Goryachev--Chaplygin Top }

In this section we shall illustrate our general procedure
by starting with the well-known Liouville integrable system 
known as the  Goryachev--Chaplygin top \cite{Whittaker, Komarov2}.
After introducing the Goryachev--Chaplygin Hamiltonian and the corresponding constant of motion, we proceed to their Eisenhart lift.
We demonstrate that the obtained four-dimensional Lorentzian spacetime, which we call the Goryachev--Chaplygin spacetime, admits a rank-3
irreducible Killing tensor. We conclude by making several comments on the quantization of the Goryachev--Chaplygin top and the corresponding results in the 
Goryachev--Chaplygin spacetime.

\subsubsection{Goryachev--Chaplygin Top}

Following Whittaker \cite{Whittaker} we consider the motion of Goryachev--Chaplygin top
as a constrained motion of a heavy top with principle moments of
inertia $A=B=4C$ and whose centre of gravity lies in the plane
determined by the two equal moments of inertia, so we start with:
\ben
L_{top}= \half ( \dot \theta ^2 + \sin ^2 \theta \dot \phi^2)   + \frac{1}{8}  (\dot \psi + \cos
\theta \dot \phi  )^2  - \alpha^2 \sin \theta \sin \psi.   
\een    
Proceeding to the Hamiltonian formulation, we find 
\ben
p_\phi = \sin ^2 \theta \dot \phi + \frac{1}{4} \cos \theta (\dot \psi
+ \cos \theta \dot \phi) \,, \qquad p_\theta = \dot \theta\,,\qquad p_\psi =
\frac{1}{4} ( \dot \psi + \cos \theta \dot \phi ) \,, 
\een
and hence the Hamiltonian is 
\bea\label{top2}
H_{top}&=& \half p_\theta ^2 + 2  p_\psi ^2 + \half
(\frac{p_\phi}{\sin \theta} - \cot \theta p_\psi ) ^2 +\alpha^2 \sin \theta \sin \psi \nonumber\\
&=&\frac{1}{2}\bigl(M_1^2+M_2^2+4M_3^2\bigr)+\alpha^2 x_2\,,
\eea
which, in notations of the appendix, is the Hamiltonian (1) considered by Komarov \cite{Komarov2}.
It is obvious that coordinate $\phi$ is cyclic and hence $p_\phi$ equals constant. 
The Hamiltonian of Goryachev--Chaplygin top is obtained if one sets $p_\phi=0$\,,  
\ben
H_{GC}= \half \bigl( \cot^2 \theta  +4\bigr )p_\psi ^2 + \half p_\theta
^2 + \alpha^2 \sin \theta \sin \psi\,. \label{top}   
\een

The Hamiltonian (\ref{top2}) has a remarkable property such that the function 
\ben\label{Ktop2}
{K}_{top}=M_3(M_1^2+M_2^2)-\alpha^2M_2x_3
\een
obeys 
\ben\label{HK}
\{H_{top}, {K_{top}}\}=\alpha^2 p_\phi M_1\,.
\een
Hence, for $p_\phi=0$, i.e. for Goryachev--Chaplygin top, (\ref{Ktop2}) is a constant of motion and reads
\ben
{K}_{GC}=p_\psi p_\theta^2+\cot^2\!\theta p_\psi^3+\alpha^2\cos\theta\bigl(\sin\psi\cot\theta p_\psi-\cos\psi p_\theta\bigr) \,.  
\een
Introducing the following functions (projections of standard functions $M_i$): 
\ben
m_1 = -\sin \psi p_\theta -\cos \psi\cot \theta p_\psi \,,\quad 
m_2 = \cos \psi p_\theta - \sin \psi \cot \theta p_\psi \,,\quad 
m_3= p_\psi \,,
\een
we may write the Goryachev--Chaplygin top Hamiltonian and the corresponding constant of motion as
\ben
H_{GC}=\frac{1}{2}\bigl(m_1^2+m_2^2+4m_3^2\bigr)+\alpha^2 x_2\,,\quad 
K_{GC}=m_3(m_1^2+m_2^2)-\alpha^2m_2x_3\,.\label{co}
\een

\subsubsection{Eisenhart lift: Goryachev--Chaplygin spacetime}
Using the results of section 2 the Hamiltonian (\ref{co}) lifts to the four-dimensional Hamiltonian
\ben
{\cal H}=  m_1^2+m_2^2+4m_3^2+2\alpha^2 p_s^2x_2+2p_sp_t\,.  
\een
This generates the geodesic flow of the four-dimensional Lorentzian
4-metric with  Killing vector fields $k=\p_t$ and $l=\p_s$, the latter
of which is lightlike and covariantly constant,
\ben
g = -2\alpha^2\sin\theta \sin\psi dt^2+2dtds+d\theta^2+\frac{d\psi^2}{\cot^2\theta+4}\,. \label{metric}
\een
The constant of motion (\ref{co}) now reads
\ben
{\cal K}=m_3(m_1^2+m_2^2)-\alpha^2 p_s^2m_2x_3
\een
and defines a rank-3 Killing tensor $K$, ${\cal K}={K}^{abc}p_a p_b p_c$, 
with non-zero contravariant components
\begin{eqnarray}\label{KT}
K^{\theta \theta \psi} = \frac{1}{3}\,,\quad 
K^{\theta s s} =- \frac{\alpha^2}{3} \cos \psi \cos \theta\,,\quad 
K^{\psi \psi \psi} = \cot^2 \theta\,,\quad 
K^{\psi s s} = \frac{\alpha^2}{3} \frac{\cos^2 \theta \sin \psi}{\sin \theta}\,,
\end{eqnarray}
together with the other components related by symmetry. One may verify
directly that $K$ satisfies the Killing equation, $\nabla^{(a}K^{bcd)}=0$, 
however, it is \emph{not} covariantly constant.

We can see in an elementary way that $K$ is not decomposable into lower rank Killing
tensors. This follows from the fact that $k$ and $l$ are the only Killing vectors 
 of the spacetime (\ref{metric}).
Suppose $K$ were decomposable,
then it would be the sum of terms of the form
\begin{equation}
K_{(1)}^{(a}K_{(2)}^{bc)}, \qquad \mathrm{or}\quad K_{(3)}^{(a}K_{(4)}^b K_{(5)}^{c)}\,,
\end{equation}
where the $K_{(i)}$ are Killing tensors. Since a rank $1$ Killing tensor
is a Killing vector, by our assumption at least one of the factors in
each term must be either $k$ or $l$. Such terms will only have
non-zero components when at least one of $a, b, c$ is either
$t$ or $s$. Since $K$ has a non-zero $\psi\psi\psi$-component, $K$ cannot be decomposed into a sum of lower rank Killing
tensors. 

One may verify that the following holds:
\begin{equation}
[ k, l] = 0\,, \quad \mathcal{L}_k K = 0\,, \quad \mathcal{L}_l K = 0\,,
\end{equation}
which implies that the associated constants of the geodesic motion are
in involution; the motion is Liouville integrable.

Let us finally mention some properties of the Goryachev--Chaplygin spacetime.
The spacetime is not Ricci flat, nor does the Ricci scalar vanish. This means that it does not admit a Killing spinor, e.g., \cite{Figueroa}. 
We also note that
\begin{equation}
R_{ab} l^b = 0,
\end{equation}
however $R_{ab}$ clearly has rank $3$ (for typical values of the
coordinates) and so $R_{ab} \neq A m_a m_b$ 
for any vector $m^a$. The Einstein tensor has non-zero components
\begin{eqnarray}
G_{tt} = \frac{-12\alpha^2(3 \cos ^4 \theta - 10 \cos^2 \theta + 6)\sin
  \theta \sin \phi}{(3 \cos^2 \theta - 4)^2}\,,\quad 
G_{ts}&=& - \frac{2(3 \cos^2 \theta + 2)}{(3 \cos^2 \theta - 4)^2}\,,
\end{eqnarray}
and obeys $G_{ab}l^a l^b = 0$\,, which is, of course, obvious 
from the equivalent result for the
Ricci tensor, together with the fact that $l$ is null.

\subsubsection{Quantum mechanics of Goryachev--Chaplygin Top}
The quantum mechanics of the Goryachev--Chaplygin top was studied by Komarov
\cite{Komarov2}. Specifically, it was shown that (\ref{HK}) admits a quantum analogue
\ben
[\hat H_{top}, {\hat K_{top}}]=-\alpha^2 J_1\partial_\phi\,,
\een
where operators $\hat H_{top}$ and ${\hat K_{top}}$ are given by 
\ben\label{HKoptop}
\hat H_{top}=\frac{1}{2}\bigl(J_1^2+J_2^2+4J_3^2\bigr)+\alpha^2 x_2\,,\quad 
\hat K_{top}=J_3(J_1^2+J_2^2)-\frac{1}{4}J_3-\frac{1}{2}\alpha^2(J_2x_3+x_3J_2)\,,\label{op}
\een
and  $J_i$ are defined in (\ref{Ji}). This means that acting on a wave function independent of $\phi$, the operators 
(\ref{HKoptop}) commute.

By employing the Eisenhart lift on these operators one finds that 
the operators 
\ben
\hat {\cal H}_{top}= J_1^2+J_2^2+4J_3^2+2\alpha^2 x_2 \partial_s^2+2\partial_s\partial_t\,,\ \  
\hat {\cal K}_{top}=J_3(J_1^2+J_2^2)-\frac{1}{4}J_3-\frac{1}{2}\alpha^2 (J_2x_3+x_3J_2)\partial_s^2\,,
\een
obey $[\hat {\cal H}_{top}, {\hat {\cal K}_{top}}]=-2\alpha^2 J_1\partial_s^2\partial_\phi$\,, and hence commute on $\phi$-independent wave function.
The former operator is precisely the standard wave operator on the Lorentzian 5-space with the metric $g_{top}$, obtained by the Eisenhart lift of $H_{top}$.
So we have, $\square_{top}\equiv g^{ab}_{top}\nabla_a\nabla_b=\hat {\cal H}_{top}$, where 
\begin{equation}\label{5Dtop}
g_{top}= 2 ds dt - 2 \alpha^2 x_2 dt^2 + (\sigma^1)^2+(\sigma^2)^2+\frac{1}{4}(\sigma^3)^2,
\end{equation}
and $\sigma^i$ are the left invariant forms on $SU(2)$ defined in (\ref{sigmai}). Moreover, the latter operator 
can be written as 
\begin{equation}
\hat{\cal K}_{top} = K_{(top)}^{abc}\nabla_a \nabla_b \nabla_c + \frac{3}{2}(\nabla_a K^{a b c}_{(top)})\nabla_b \nabla_c - \frac{1}{2}K_{(top) a}{}^{a b} \nabla_b\,, \label{Fkeqn}
\end{equation}
where $K_{(top)}$ is a symmetric rank-3 tensor. Introducing the basis 
\ben\label{Li}
L_s =\partial_s\,,\quad 
L_t =\partial_t\,,\quad L_i = J_i\,,
\een
one finds that non-vanishing contravariant components of $K_{(top)}$ are 
\begin{equation}\label{K5D}
K^{s s 2}_{(top)} = -2 \alpha^2 x_3/3\,, \quad K^{1 1 3}_{(top)} = K^{2 2 3}_{(top)} = 2/3\,,
\end{equation}
and that the tensor satisfies $\nabla^{(a}K_{(top)}{}^{bcd)}=-\alpha^2L_s^{(a}L_s^b(\partial_\phi)^cL_1^{d)}$\,. Hence, if we restrict to geodesic motion 
on 5-space with metric $g_{top}$ such that $p_\phi$ vanishes, $K_{(top)}^{abc}p_ap_bp_c$ defines a constant of motion.

One might wonder whether it is possible to directly carry over the quantization to the  Goryachev--Chaplygin four-dimensional spacetime discussed in the previous subsection.
The `naive quantization' of (\ref{co}) gives 
\ben
\hat H_{GC}=\frac{1}{2}\bigl(j_1^2+j_2^2+4j_3^2\bigr)+\alpha^2 x_2\,,\quad 
\hat K_{GC}=j_3(j_1^2+j_2^2)-\frac{1}{4}j_3-\frac{1}{2}\alpha^2(j_2x_3+x_3j_2)\,,\label{op}
\een
where we have defined the operators (projections of $J_i$)
\ben
j_1 = -\sin \psi \partial_\theta -\cos \psi\cot \theta \partial_\psi \,,\quad 
j_2 = \cos \psi \partial_\theta - \sin \psi \cot \theta \partial_\psi \,,\quad 
j_3= \partial_\psi \,.
\een
By lifting the operators (\ref{op}), one finds
\ben\label{ops2}
\hat {\cal H}= j_1^2+j_2^2+4j_3^2+2\alpha^2 x_2 \partial_s^2+2\partial_s\partial_t\,,\ \ 
\hat {\cal K}=j_3(j_1^2+j_2^2)-\frac{1}{4}j_3-\frac{1}{2}\alpha^2 (j_2x_3+x_3j_2)\partial_s^2\,.
\een
It is easy to verify that $[\hat {\cal H}, \hat {\cal K}]=0$. However, the operator $\hat {\cal H}$ is not a standard (geometrical) wave operator on the 
Goryachev--Chaplygin spacetime. In fact, one finds
\ben
\square\equiv g^{ab}\nabla_a\nabla_b=\hat {\cal H}-\frac{3\cot \theta}{4+\cot^2\!\theta}\partial_\theta\,.
\een
It is an interesting question whether the operators (\ref{op}) provide the `correct quantization' of the Goryachev--Chaplygin top, in which case the operators 
(\ref{ops2}) are `preferred operators' in the Goryachev--Chaplygin
spacetime, or whether some alternative quantization is more
appropriate. We leave this problem for the future. We also remark that
we were not able to find an operator linear in the Killing tensor $K$,
(\ref{KT}), which commutes with the wave operator $\square$
associated with the Goryachev--Chaplygin metric (\ref{metric}).

\subsection{Kovalevskaya's Spacetime: Quartic Killing Tensor}
In this case one considers a heavy top with principle moments of 
inertia $A=B=2C$ 
whose centre of gravity  lies in the plane determined by the two equal
moments of inertia. The Lagrangian is 
\ben
L_{K}= \frac{1}{2}(\dot \theta ^2 + \sin ^2 \theta \dot \phi^2)+\frac{1}{4}(\dot \psi + \cos
\theta \dot \phi)^2   -\alpha^2 \sin \theta \cos \psi \,.
\een 
 Clearly $\phi$ is ignorable and the Hamiltonian 
\bea
H_{K}&=& \half \left( p_\theta ^2 +(\frac{p_\phi}{\sin\theta}-\cot \theta p_\psi)^2 
 + 2 p_\psi ^2\right ) + \alpha^2 \sin \theta \cos \psi \nonumber\\
 &=& \half \left( M_1^2+M_2^2+2M_3^2\right)+ \alpha^2 x_1
\eea
is constant.
Kovalevskaya found another constant \cite{Whittaker, Borisov} which reads
\bea
K_{K}&=& \left( p_\theta ^2 +(\frac{p_\phi}{\sin\theta}-\cot \theta p_\psi)^2\right)^2
+ 4 \alpha^4 \sin
^2 \theta - 2 \alpha^2 \sin \theta \left(e^{i\psi}(\frac{p_\phi}{\sin\theta}- \cot \theta p_\psi + i p_\theta
)^2 + c.c. \right) \nonumber\\
 &=& (M_1^2+M_2^2)^2+ 4 \alpha^4 (x_1^2+x_2^2)
- 4 \alpha^2 \left[x_1(M_1^2-M_2^2)+2x_2M_1M_2\right] \,.
\eea
This will lift to give a quartic Killing tensor.

In order to get a four-dimensional spacetime we 
perform again the reduction along the $\phi$-direction. So we consider
\bea
 H\!\!&=&\!\!\half \left( m_1^2+m_2^2+2m_3^2\right)+ \alpha^2 x_1\,,\nonumber\\
 K \!\!&=&\!\!(m_1^2+m_2^2)^2+ 4 \alpha^4 (x_1^2+x_2^2)
- 4 \alpha^2 \left[x_1(m_1^2-m_2^2)+2x_2m_1m_2\right]\,.
\eea
The Hamiltonian lifts to 
\ben
{\cal H}=m_1^2+m_2^2+2m_3^2+ 2\alpha^2 p_s^2x_1+2p_sp_t\,,
\een
which generates geodesic flow of the Lorenzian 4-metric
\begin{equation}
g = - 2 \alpha^2 \sin\theta \cos\psi dt^2 +  2 ds dt+d\theta^2+\frac{d\psi^2}{\cot^2\!\theta+2}\,,
\end{equation}
admitting the rank-4 irreducible tensor $K$, given by 
\bea
K^{\theta\theta\theta\theta}\!\!\!&=&\!\!\!1\,,\ \ K^{\theta\theta \psi\psi}=\frac{1}{3}\cot^2\!\theta\,,\ \ 
K^{ss\theta\theta}=\frac{2}{3}\alpha^2\sin\theta\cos\psi\,,
\ \  
K^{\psi\psi\psi\psi}=\cot^4\!\theta\,,\nonumber\\
K^{ss\theta\psi}\!\!\!&=&\!\!\!-\frac{2}{3}\alpha^2\cos\theta\sin\psi\,,\ \  
K^{ss\psi\psi}=-\frac{2}{3}\alpha^2\cos\psi\cos\theta \cot\theta\,,\ \ 
K^{ssss}=4\alpha^4\sin\theta^2\,.
\eea
Properties of the Kovalevskaya spacetime are very similar to properties of the Goryachev--Chaplygin spacetime.
In particular, the spacetime admits a covariantly constant null Killing vector $l=\partial_s$, it is not Ricci flat, and does not admit a Killing spinor.
We also have that $\square\neq \hat {\cal H}$, with the latter obtained by a naive quantization described in previous section. 

One can again consider a 5D spacetime instead,
\begin{equation}
g_{K}= - 2 \alpha^2 \sin\theta \cos\psi dt^2 +  2 ds dt + (\sigma^1)^2+(\sigma^2)^2 + \frac{1}{2}(\sigma^3)^2\,,
\end{equation}
where one has \cite{Komarov1}
\bea
\square_{K}&=&g^{ab}_{K}\nabla_a\nabla_b=\hat {\cal H}_{K}=J_1^2+J_2^2+2J_3^2+ 2\alpha^2 x_1\partial_s^2+2\partial_s\partial_t\,,\nonumber\\
\hat {\cal K}_K&=&\frac{1}{2}(K_+K_-+K_-K_+)-2(J_+J_-+J_-J_+)\,,
\eea
where $J_\pm=J_1\pm iJ_2$, $K_\pm=J_\pm^2-2\alpha^2x_\pm\partial_s^2$ and $x_\pm=x_1\pm ix_2$.
In this case $\hat {\cal K}_{K}$ is a real symmetry of the wave operator, $[\square_K, \hat {\cal K}_{K}]=0$. It is related to the five-dimensional rank-4 irreducible Killing tensor $K_{(K)}$ as    
\begin{eqnarray}
\hat{\cal K}_K &=& K^{abcd}_{(K)}\nabla_a \nabla_b \nabla_c \nabla_d + 2 (\nabla_a
K_{(K)}^{abcd}) \nabla_b \nabla_c \nabla_d + 3 (\nabla_a \nabla_b
K_{(K)}^{abcd})\nabla_c \nabla_d \nonumber \\
&& - 2 K_{(K)}^{abc}{}_c\nabla_a \nabla_b -
\frac{3}{4}K_{(K)}^{ab}{}_{ab} L_3^c L_3 ^d \nabla_c \nabla_d\,,
\end{eqnarray}
where in the basis (\ref{Li}) the components of the Killing tensor $K_{(K)}$ are written as 
\begin{eqnarray}
K^{ssss}_{(K)} &=& 4\alpha^4(x_1^2+x_2^2)\,, \qquad K^{ss11}_{(K)}=-K^{ss22}_{(K)} = -2 \alpha^2 x_1/3\,, \nonumber \\
K^{ss12}_{(K)} &=& -2 \alpha^2 x_2/3\,, \qquad K^{1111}_{(K)} = 3 K^{1122}_{(K)} = K^{2222}_{(K)} = 1\,.
\end{eqnarray}

\section{Superintegrable systems in $E^2$: SUSY plane waves}
In this section we consider Hamiltonians of the form 
\bea
H = \frac{1}{2}(p_x^2+p_y^2)+V(x,y) ~. \label{KMPH}
\eea
For some choices of the potential $V$ this Hamiltonian is superintegrable, e.g., \cite{Kalnins:2009} and references therein. The Hamiltonian (\ref{KMPH}) lifts to
\bea
{\cal H} = p_x^2+p_y^2+2V(x,y)p_s^2+2p_sp_t\,, 
\eea
which generates geodesic flows of Lorentzian 4-metric
\bea
g = dx^2+dy^2-2V(x,y)dt^2+2dtds ~. \label{KMPmet}
\eea
In quantum mechanics, one has the quantized Hamiltonian
\bea
\hat{{\cal H}} = \partial_x^2+\partial_y^2+2V(x,y)\partial_s^2+2\partial_s\partial_t
\eea
and this coincides with the Laplacian of the metric (\ref{KMPmet}), i.e., one has $\square\equiv \nabla_ag^{ab}\nabla_b=\hat{{\cal H}} ~.$

Let us mention some basic properties of the spacetime (\ref{KMPmet}).
The Ricci curvature has only $tt$-component,
\bea
R_{tt} = (\partial_x^2+\partial_y^2)V ~,
\eea
and the scalar curvature vanishes, $R=0$. Hence $G_{ab}=R_{ab}$ and 
\bea
R_{ab}l^b = 0 ~,
\eea
where $l\equiv\partial/\partial s$ is a covariantly constant null Killing vector. 
Since the ``transverse'' $x$-$y$ space is flat the metric  (\ref{KMPmet})
admits a covariantly constant 
spinor field $\epsilon$ such that  
\ben
\bar \epsilon \gamma ^a \epsilon    =  l^a = (\partial_s)^a   
\een
 and hence a covariantly constant null 2-form
\ben
\ell^{ab }= \bar \epsilon \gamma ^{[ab]} \epsilon
\een   
such that $\ell^{ab } l_b=0\,.$

There are many examples of interesting (superintegrable) systems of the type (\ref{KMPH}) which give rise to higher-rank Killing tensors
and non-trivial Schouten--Nijenhuis brackets. We refer the reader to recent paper by Kalnins {\em et al.} \cite{Kalnins:2009} and references therein as well as to 
Chapter 4.4 in \cite{Pettini}. To illustrate the theory we give the following recent example:

\subsection{Post--Winternitz example}
In \cite{Post}, Post and Winternitz give a (Hamilton--Jacobi non-separable) 
classical super-integrable example 
of the form (\ref{KMPH}) with the potential  
\ben
V=\frac{\alpha y }{x^{\frac{2}{3} }} \,,
\een
such that 
\bea
X &=&3p_x^2 p_y + 2 p_y^3 + 9 \alpha
 x^{\frac{1}{3}}p_x  + \frac{6 \alpha y p_y}{x^{\frac{2}{3}}} \,, \\  
 Y &=& {p_x}  ^4 +
\frac{4 \alpha y p_x  ^2 }{x^{\frac{2}{3}} } - 
12\alpha x^{\frac{1}{3}} p_x p_y - 
\frac{2 \alpha ^2 (9x^2-2y^2)}{x^{\frac{4}{3}}} \,,
\eea
both Poisson commute with $H$ and satisfy the Heisenberg algebra 
 \ben
\{ X,Y \} = 108 \alpha^3  \,.  \label{Heisenberg} 
\een
The spacetime reads 
\ben
g = 2dsdt -\frac{2y}{x^{\frac{2}{3}}}dt^2 + dx^2+dy^2 \,.
\een
The constants $X,Y$ are lifted and give  
\ben
\{ {\cal X},{\cal Y}  \} = 108 \alpha^3 p_s^6  \,. 
\een
Thus, consistent with previous cases (\cite{GDH} and references therein),
the central  element in the Heisenberg algebra  (\ref{Heisenberg}) 
may be interpreted as the (sixth power of) a null translation.

The spacetime admits rank-3 and rank-4 Killing tensors. Their components $X^{abc}$ and $Y^{abcd}$ can be read of from 
\bea
\mathcal{X} \!\!\!&=&\!\!\!X^{abc}p_a p_b p_c=3p_x^2 p_y + 2 p_y^3 + 9 \alpha
 x^{\frac{1}{3}}p_x p_s^2  + \frac{6 \alpha y p_y p_s^2}{x^{\frac{2}{3}}} \,, \\  
 \mathcal{Y} \!\!\!&=&\!\!\!Y^{a b c d} p_a p_b 
p_c p_d = {p_x}  ^4 +
\frac{4 \alpha y p_x  ^2 p_s^2}{x^{\frac{2}{3}} } - 
12\alpha x^{\frac{1}{3}} p_x p_y p_s^2 - 
\frac{2 \alpha ^2 (9x^2-2y^2)}{x^{\frac{4}{3}}}p_s^4 \,.
\eea
Since $l_a dx^a = dt$, we have  
\ben
l_a X^{abc} =0 = l_a  Y^{abcd}\,.
\een

Post and Winternitz have provided a quantization of their model.
Thus if $[x, p_x]= i \hbar$  etc, then all products are replaced by half
their anti-commutator and in addition one must subtract  $\frac{5 \hbar
  ^2}{72 x^2}$ from the 
expression for $H$ and add  
$\frac{25 \hbar ^4}{1296 x^4}$ to the expression for $Y$ .

\section{Conclusions}

In this paper we have shown that by applying Eisenhart's lightlike
lift to dynamical systems admitting constants of the motion of degree
greater than two in momenta, one may obtain spacetimes admitting
Killing tensors of higher rank than two. Our examples by no means
exhaust the possibilities. In
\cite{Borisov,Komarov1, Kalnins:2009, Komarov3, Komarov4,Komarov5,Dullin, Valent} more
complicated examples are given, but our examples illustrate the point
we wish to make.

In some cases we find the
Poisson--Schouten--Nijenhuis algebra to be non-trivial. We have also
constructed differential operators which realize the classical algebra
as $\hbar\to 0$. In some, but not all, cases the Hamiltonian
corresponds to the Laplace or wave operator. In general the wave
operator must be augmented by quantum corrections which are not always
expressible in purely geometric terms. The higher rank conserved
quantities also receive quantum corrections not expressible solely in
terms of the Killing tensor. In some ways this is one of the
most interesting of our findings and is certainly worthy of further study.

\section*{Acknowledgments}
We would like to thank C. Duval, P. Horvathy and G. Valent for comments on the draft and M. Dunajski for helpful discussions. The work of T.H. is supported by the JSPS Institutional Program for
Young Researcher Overseas Visits ``Promoting international young
researchers in mathematics and mathematical sciences led by OCAMI''.
He also would like to thank DAMTP, University of Cambridge, for the hospitality.
D.K. acknowledges the Herchel Smith Postdoctoral Fellowship at the
University of Cambridge.

\appendix

\section{Conventions and Euclidean Group notation} 
To fix the conventions for forms on $SU(2)$, we take the following basis for left-invariant forms:
\ben\label{sigmai}
\sigma^1 = \sin \theta \cos \psi d \phi - \sin \psi d\theta\,,\quad 
\sigma^2 = \sin \theta \sin \psi d \phi + \cos \psi d\theta\,,\quad 
\sigma^3 = d\psi + \cos \theta d\phi\,,
\een
which obey the relations
\begin{eqnarray}
d \sigma^i = -\frac{1}{2} \epsilon_{ijk} \sigma^j \wedge \sigma^k\,.
\end{eqnarray}
The dual vector fields are
\ben\label{Ji}
J_1 = -\sin\psi \partial_\theta + \frac{\cos\psi}{\sin\theta}\partial_\phi - \cot\theta \cos \psi\partial_\psi\,,\ 
J_2 = \cos\psi \partial_\theta + \frac{\sin\psi}{\sin\theta}\partial_\phi - \cot\theta \sin \psi\partial_\psi\,,\   
J_3 = \partial_\psi\,,
\een
and satisfy the algebra:
\begin{equation}
\left[ J_i, J_j\right] = - \epsilon_{ijk}J_k\,.
\end{equation}
Defining the functions
\ben\label{xi}
x_1 = \sin \theta \cos \psi\,,\quad 
x_2 = \sin \theta \sin \psi\,,\quad 
x_3 = \cos \theta\,,
\een
we have the additional relations
\begin{equation}
\left[ J_i, x_j\right] = - \epsilon_{ijk}x_k\,,
\end{equation}
where we interpret the functions $x_i$ as operators on functions, acting by multiplication.

Both the Goryachev--Chaplygin and the Kovalevskaya tops
discussed in the main text are examples
of tops whose centre of gravity does not coincide with
the pivot point. They admit a  description
in terms of the Lie algebra  of the Euclidean group $E(3)$
and since this is used in some of the literature, e.g.  \cite{Komarov2,Borisov,
  Komarov1,  Komarov3, Komarov4, Bolsinov}, 
we give it here. 
 
If $\bM$ is the angular momentum of the top
one has, in the rotating frame
\bea
\dot \bM + \bom \times \bM &=& -mg \bx_0 \times \bx\,, \nonumber\\ 
\dot \bk + \omega \bx &=&0 \label{sytem} \,,
\eea
where $\bx$ is  unit vector which is constant  in the inertial frame (the constancy of $|\bx|$ is a consequence of these 
equations of motion)
and points in the opposition  direction to the local direction of gravity
and $\bx _0$ is a constant  vector in the rotating from
which gives the centre of gravity. An alternative interpretation,
used in analyzing the {\it Stark effect}, 
is that $\bx _0$ is the electric dipole moment and and $mg \bx$
is in the direction of the  applied electric field.    
The system of equations admits three  constants of the motion
\ben
\bx \cdot \bx\,,\quad 
\bx \cdot \bM\,,\quad 
\half \bom \cdot \bM + mg \bx_0 \cdot \bx \,.  
\een 
Choosing coordinates  such that the   centre of mass relative to the pivot
(normalized to unit length) are given by (\ref{xi}), 
we find that the  potential energy of the top is given by  
\ben
V= mg ( x_0  \sin \theta \cos \psi + y_0 \sin \theta \sin \psi +
z_0 \cos \theta )     \,,
\een
and one may construct a Lagrangian on $TSO(3)$ and  a  Hamiltonian
on $T^\star SO(3)$ which depend on the principle moments of 
inertia $(A,B,C)$. For the Goryachev--Chaplygin top
we have $A=B=4C$, and the centre of gravity lies in the 
plane defined by the two principal axes with equal moments of inertia. 

The moment maps  for left actions of rotations
\ben\label{Mi}
M_1 = -\sin \psi p_\theta + \frac{\cos \psi}{\sin\theta}p_\phi-\cos\psi\cot \theta p_\psi \,,\  
M_2 =\cos \psi p_\theta + \frac{\sin \psi}{\sin\theta}p_\phi-\sin\psi\cot \theta p_\psi \,,\  
M_3= p_\psi \,.
\een
The Poisson algebra of $\bf M$ and  $\bf x$ then turns out to be that
of the Euclidean group $\frak{e}(3)$. 
Thus the system of equations (\ref{sytem}) may also be interpreted as 
a Hamiltonian system  moving
on $\frak{e} ^\star(3)$ the dual of the Lie algebra $\frak{e}(3)$.
As a consequence one has an isomorphism with the problem of a rigid body
moving in a fluid.  However it should be noted that  the 
latter has phase space $T^ \star E(3)$
which is 12-dimensional while the top has phase space
has phase space $T^\star (SO(3))$ which is 6-dimensional.
As pointed out in \cite{Bolsinov} if one imposes the constraints
$\bx \cdot \bx =1\,, \bM \cdot \bx =0$, one gets the standard
symplectic structure on $T^\star S^2$.

\end{document}